\documentclass[journal]{IEEEtran}
\usepackage[table]{xcolor}
\usepackage[english]{babel}
\usepackage[utf8]{inputenc}
\usepackage{eurosym}
\usepackage{float}
\usepackage{flushend}
\usepackage{setspace}
\usepackage{amssymb}
\usepackage{tikz}
\usetikzlibrary{fit}
\usetikzlibrary{backgrounds}
\usetikzlibrary{positioning,fit,calc}
\usepackage[]{tabularx}
\usepackage{circuitikz}

\usepackage{booktabs,multirow}

\usepackage{moreverb}
\usepackage{enumerate}
\usepackage{textcomp}
\usepackage{graphicx}
\usepackage{amssymb}
\usepackage{amsthm}
\usepackage{amsmath}
\usepackage{multirow}
\usepackage{array}
\usepackage{bm}
\usetikzlibrary{arrows}
\usepackage{enumitem}
\usepackage{arydshln}
\usepackage[noadjust]{cite}

\allowdisplaybreaks

\begin{document}
	\title{Geometric Algebra Power Theory (GAPoT): Revisiting Apparent Power under Non-Sinusoidal Conditions}
	\vspace{-0.1cm}
	\author{Francisco Gil Montoya, Alfredo Alcayde, Francisco Arrabal-Campos, Raul Ba\~{n}os, \\ and   Javier Rold\'{a}n-P\'{e}rez, {\em Member, IEEE}
    }	
	\maketitle
	\vspace{-0.9cm}
\begin{abstract} 
Traditional power theories and one of their most important concepts --apparent power-- are still a source of debate and, as shown in the literature, they present several flaws that misinterpret the power-transfer phenomena under distorted grid conditions. In recent years, advanced mathematical tools such as geometric algebra (GA) have been applied to address these issues. However, the application of GA to electrical circuits requires more consensus, improvements and refinement.
In this paper, power theories based on GA are revisited. Several drawbacks and inconsistencies of previous works are identified and modifications to the so-called geometric algebra power theory (GAPoT) are presented. This theory takes into account power components generated by cross-products between current and voltage harmonics in the frequency domain. Compared to other theories based on GA, it is compatible with the traditional definition of apparent power calculated as the product of RMS voltage and current. 
Also, mathematical developments are done in a multi-dimensional Euclidean space where the energy conservation principle is satisfied.
The paper includes a basic example and experimental results in which measurements from a utility supply are analysed. Finally, suggestions for the extension to three-phase systems are drawn.
\end{abstract}
\vspace{-0.15cm}
\begin{IEEEkeywords}
Geometric algebra, non-sinusoidal power, Clifford algebra, power theory.
\end{IEEEkeywords} 
\vspace{-0.25cm}
\section{Introduction}
Full understanding of power flows in electrical systems has been a topic of interest during the last century.
The most relevant efforts have been done in the frequency domain for systems operating in steady state~\cite{czarnecki2005currents}, and in the time domain by using both instantaneous and averaged approaches~\cite{akagi2017instantaneous,staudt2008fryze}.
Outcomes from these studies are sometimes inconsistent and even contradictory.
%
For example, the well-known {\em instantaneous power theory} can yield to incoherent results under specific conditions~\cite{czarnecki2004some}. Similar conflicting results have been found for well-established regulations such as the IEEE standard 1459~\cite{castro2012ieee}.
Traditional techniques that are commonly applied for analysing power flows are based on linear-algebra tools such as complex numbers, matrices, tensors, etc, and they are proven to be useful from the application point of view. 
However, none of them provide a clear overview of power flows under disported and unbalanced grid conditions and this point is still an open discussion~\cite{petroianu2015geometric}.

Geometric algebra (GA) is a mathematical tool developed by Clifford and Grassmann by the end of the XIX century that has been refined by Hestenes in the last decade~\cite{hestenes2012clifford}.
This tool has brought new possibilities to the physics field such as producing compact and generalised formulations~\cite{hestenes2012new}. Also, it can be easily used to manipulate integral and differential equations in multi-component systems~\cite{dorst2010geometric,chappell2014geometric}.
%
%
%
Even though GA is not widely known by the scientific community, it has a great potential and has attracted interest in recent publications~\cite{petroianu2015geometric}.
%
%
GA has already been introduced to redefine the apparent power as the geometric product between voltage and current, what is commonly written as $\bm{M}$~\cite{menti2007geometric,castilla2008clifford,lev2009geometric,montoya2018power}.
%
Compared to the traditional definition of apparent power ($S=VI$), the use of $\bm{M}$ has several advantages. A relevant one is that $\bm{M}$ is conservative in spite of $S$ and this is of interest for its application in distorted environments~\cite{castro2016m}.
%

In this paper, GA power theories proposed by different authors are reviewed in order to analyse some of the inconsistencies raised so far, while additional ones not yet found in the literature are also discussed~\cite{castilla2008geometric,castro2012ieee,menti2007geometric}. 
%
%
%
Then, the GA power theory is redefined in order to solve these issues. The proposed theory can be used to resolve electrical circuits, is compatible with the traditional definition of apparent power and provides a definition for its components that fulfils the principle of energy conservation.
%
%
Numerical and experimental results are included in order to validate the main contributions of this work.
A brief introduction to GA and its terminology is included in order to make the paper self-contained.
%
\vspace{-0.15cm}
\section{Geometric Algebra for Power Flow Analysis}
%
%
The geometric product was introduced by Clifford by the end of the XIX century, and it includes the external and internal products of two vectors, namely 
%
$\bm{a}=\alpha_1\bm{\sigma_1}+\alpha_2\bm{\sigma_2}$ and $\bm{b}=\beta_1\bm{\sigma_1}+\beta_2\bm{\sigma_2} \in \mathbb{R}^2$~\cite{hestenes2012clifford}. The internal product can be calculated as follows:
\begin{equation}
\bm{a}\cdot\bm{b}=\|\bm{a}\|\|\bm{b}\|\cos{\varphi}=\sum{\alpha_i\beta_i}
\label{eq:inner}
\end{equation}
\noindent while the external product is:
\begin{equation}
\bm{a}\wedge\bm{b}=\|\bm{a}\|\|\bm{b}\|\sin{\varphi}\ \bm{\sigma_1\sigma_2}
\label{eq:outer}
\end{equation}
%
%
This operation does not exist in traditional linear algebra and its result is not a scalar nor a vector, but a new entity that is commonly known as~\textit{bivector}~\cite{hestenes2012clifford}. 
%
Bivectors play a key role in calculations related to non-active power, as will be shown later.
The external product is anticommutative, i.e., $ \bm{a} \wedge \bm{b} = -\bm{b} \wedge \bm{a}$. 

The fundamental operation in GA is the geometric product:
\begin{equation}
\begin{aligned}
\bm{M}=\bm{a}\bm{b}&=(\alpha_1\bm{\sigma_1}+\alpha_2\bm{\sigma_2})(\beta_1\bm{\sigma_1}+\beta_2\bm{\sigma_2})\\
&=(\alpha_1\beta_1+\alpha_2\beta_2)+(\alpha_1\beta_2-\alpha_2\beta_1)\bm{\sigma_1\sigma_2}
\end{aligned}
	\label{eq:geom_product}
\end{equation}
\noindent where $\bm{M}$ consists of two elements. As these elements are of a different nature, $\bm{M}$ is commonly referred to as {\em multivector}. The operator $\langle \cdot \rangle_k$ refers to $k$-grade component of a multivector. In (\ref{eq:geom_product}), the term $ \langle\bm{M}\rangle_0$ is is a scalar, while the term $\langle\bm{M}\rangle_2$ is a bivector.  Multivectors are classified according to their degree:
%
scalars have degree zero, vectors one, bivectors two, etc.
%
The norm of a multivector is:
%
\begin{equation}
	\|\bm{M}\|=\sqrt{\langle\bm{M}^{\dagger}\bm{M}\rangle_0}
\end{equation} 
\noindent where $\bm{M}^{\dagger}$ is the reverse of $\bm{M}$ (see~\cite{hestenes2012clifford} for details).

Considering a single-phase system operating under perfect sinusoidal conditions, it is possible to select an orthonormal basis such as
$\bm{\sigma}=\{\bm{\sigma}_1 \rightarrow\sqrt{2}\cos\omega t, \bm{\sigma}_2 \rightarrow\sqrt{2}\sin \omega t\}$. Therefore, voltages and currents are transformed as follows:
%
\begin{equation}
\begin{aligned}
	u(t) &\longrightarrow &\bm{u}&=\alpha_1\bm{\sigma_1}+\alpha_2\bm{\sigma_2} \\
	i(t) &\longrightarrow &\bm{i}&=\beta_1\bm{\sigma_1}+\beta_2\bm{\sigma_2}
\end{aligned}
\end{equation}
The geometric product defined in (\ref{eq:geom_product}) can be used to calculate the {\em geometric apparent power:}
%
\begin{equation}
	\bm{M}=\bm{u}\bm{i}=\underbrace{(\alpha_1\beta_1+\alpha_2\beta_2)}_{P}+\underbrace{(\alpha_1\beta_2-\alpha_2\beta_1)}_{Q}\bm{\sigma_1\sigma_2} 
\end{equation}
This expression consists of two terms that can be clearly identified: $P$ is a scalar and $Q$ is a bivector.
This result will be extended to non-sinusoidal conditions in  later sections. 

The geometric apparent power fulfils:
\begin{equation}
	\|\bm{M}\|^2=\langle\bm{M}\rangle_0^2+\langle\bm{M}\rangle_2^2=P^2+Q^2=\|\bm{u}\|^2\|\bm{i}\|^2
\end{equation} 
%
\section{GA-Based Power Theories: Overview}
\label{sec:overview}
In this section, the main power theories based on GA are critically discussed so that the main contributions of this paper can be better understood.
%

\begin{table*}[ht]
	\centering
	\begin{tabular}{@{}ll@{}}
		\toprule
		\multicolumn{1}{c}{\textbf{Author}}                    & \multicolumn{1}{c}{\textbf{Contribution}}                                   \\ \midrule
		\textbf{Menti}                        & Pioneer definition of geometric power in electrical circuits.                              \\
		\textbf{Castilla-Bravo}                  & Generalized complex geometric algebra. Vector basis and complex-bivector power.                                            \\
		\textbf{Lev-Ari}                & Time domain and multiphase power introduction.                                  \\
		\textbf{Castro-N\'{u}ñez}                                 & Circuit analysis and definition of geometric impedance. Conservative geometric power demonstration.                           \\
		\textbf{Montoya}                                     & Corrections on definition of power components. Optimal current decomposition. Interharmonics.                                              \\
		\bottomrule
	\end{tabular}
	\vspace{+0.1cm}
	\caption{Main contributors to GA-based power theories.}
	\vspace{-0.6cm}
	\label{tab:power_contributions}
\end{table*}

\begin{itemize}
	\item \textbf{Menti}. This theory was developed by Anthoula Menti {\em et al.} in 2007~\cite{menti2007geometric}. This was the first application of GA to electrical circuits. The apparent power multivector was defined by multiplying the voltage and current in the geometric domain:
	%
	\begin{equation*}
		\bm{S}=\bm{u}\bm{i}=\bm{u}\cdot\bm{i}+\bm{u}\wedge\bm{i}=\left\langle \bm{S}\right\rangle_0
+\left\langle \bm{S}\right\rangle_2
	\end{equation*}
	The scalar part matches the active power $P$, while the bivector part represents power components with a mean value equal to zero.
	It was demonstrated that the latter holds for both sinusoidal and non-sinusoidal conditions, in steady-state.
	%
	It was also demonstrated through examples that this theory can be applied to electrical circuits already studied in the literature, in which the components of the traditional apparent power were not distinguishable. The reason is that bivector terms provide sense and direction, while the traditional definition of apparent power based on complex numbers does not.
	Unfortunately, the theory did not establish a general framework for the resolution of electrical circuits under distorted conditions.
	Also, the proposal was not applied to decompose currents (for non-linear load compensation, for example), and it was not extended to multi-phase systems.
	%
	\item \textbf{Castilla-Bravo}. This theory was developed by Castilla and Bravo in 2008~\cite{castilla2008clifford}. Authors introduced the concept of {\em generalised complex geometric algebra}. Vector-phasors were defined for both voltage and current:
	%
	\begin{flalign*}
		\widetilde{\bm{U}}_p&=U_pe^{j\alpha_p}\bm{\sigma}_p=\bar{U}_p\bm{\sigma}_p \\
		\widetilde{\bm{I}}_q&=I_qe^{j\beta_q}\bm{\sigma}_q=\bar{I}_q\bm{\sigma}_q
	\end{flalign*}
	Geometric power results from multiplying the harmonic voltage and conjugated harmonic current vector-phasors:
	\begin{equation*}
		\widetilde{\bm{S}}=\sum_{\substack{p \in N \cup L \\ q \in N \cup M}}  	\widetilde{\bm{U}}_p \widetilde{\bm{I}}^{*}_q = \widetilde{P}+j\widetilde{Q}+\widetilde{D}
	\end{equation*}
	This proposal is able to capture the multicomponent nature of apparent power through the so-called {\em complex-scalar} $\widetilde{P}+j\widetilde{Q}$ and the {\em complex bivector} $\widetilde{D}$.
	%
	However, this formulation requires the use of complex numbers, which could have been avoided by using appropriate bivectors~\cite{hestenes2012new}.
	%
	Also, only definitions of powers were presented and it was not extended to multi-phase systems.
	
	\item \textbf{Lev-Ari}. This theory was developed by Lev Ari~\cite{lev2009geometric,lev2009instantaneous}, and it was the first application of GA to multi-phase systems in the time domain.
	%
	However, this work does not contain examples nor fundamentals for load compensation. Also, practical aspects required to solve electrical circuits were not explained.
	\item \textbf{Castro-N\'{u}\~{n}ez}. 
	This theory was developed by Castro N\'{u}\~{n}ez in the year 2010~\cite{castro2010use}, and then extended and refined in several later works~\cite{castro2012advantages,castro2012ieee,castro2016m,castro2019theorems}.
	%
	A relevant contribution of this work consists on the resolution of electrical circuits by using GA (without requiring complex numbers).
	%
	%
	Also, a multivector called {\em geometric apparent power} that is conservative and fulfils the Tellegen theorem is defined~\cite{castro2019theorems}.
	%
	As in Menti and Castilla-Bravo proposals, the results are presented only for single-phase systems.
	Another contribution is the definition of a transformation based on $k$ vectors that form an orthonormal base:
	%
	\begin{equation}
	\arraycolsep=1.0pt\def\arraystretch{1}
	\begin{array}{lcl}
	\varphi_{c1}(t) = \sqrt{2}\cos{\omega t} \quad &\longleftrightarrow  & \quad  \phantom{-}\bm{\sigma_1} \\
	\varphi_{s1}(t) = \sqrt{2}\sin{\omega t} \quad & \longleftrightarrow & \quad  \bm{-\sigma_2} \\
	& \vdots\\
	\varphi_{cn}(t) = \sqrt{2}\cos n\omega t \quad & \longleftrightarrow & \quad  \boldsymbol{\bigwedge\limits_{i=2}^{n+1} \sigma_i} \\
	\varphi_{sn}(t) = \sqrt{2}\sin n\omega t \quad & \longleftrightarrow & \quad  \boldsymbol{\bigwedge\limits_{\substack{i=1\\i \ne 2}}^{n+1} \sigma_i} \\
	\end{array}
	\label{eq:milton_transformation}
	\end{equation}
	However, this basis presents some drawbacks.
	%
	The main one is the definition of the geometric power~\cite{montoya2020geometric}. 
	In particular, active power calculations do not match with those obtained by using classical theories. Therefore, authors needed to include an ad-hoc corrective coefficient~\cite{castro2012ieee}.
	%
	Also, in the current version of the theory, it is not possible to establish optimal current compensation since current decomposition did not consider the minimal active current proposed by Fryze and supported by other authors \cite{czarnecki2005currents,staudt2008fryze,depenbrock2003theoretical}.
	%
	Finally, the definition of the geometric power does not follow the traditional expression $\|S\|=\|U\|\|I\|$, due to the transformation presented in (\ref{eq:milton_transformation}).
	
	\item \textbf{Montoya}. 
	This framework was proposed by Montoya {\em et al.} \cite{montoya2020geometric},
	and it is an extension of Menti and Castro-N\'{u}ñez theories~\cite{menti2007geometric,castro2012ieee}. It establishes a general framework for power calculations in the frequency domain~\cite{montoya2020geometrictime}.
	%
	Since it is the most recent work, it provides solutions to some problems detected so far in other proposals and the formulation is more compact and efficient. 
	%
	The optimal current for load compensation  is presented, 
	what is helpful for power quality applications. 
	Inter- and sub-harmonics can be easily modelled~\cite{MONTOYA2019486}. 
	Also, the definition of apparent power is valid for distorted and non-distorted voltages and currents~\cite{montoya2018power}.
	%
	However, this framework is based on the use of $k$-vectors. Therefore, drawbacks related to the non-standardised definition of apparent power and the fulfilment of the energy conservation principle are inherit from previous theories. Also, harmonic power components cannot be easily decomposed since this would require inverting geometric vectors.
\end{itemize}	
The most relevant contributions to power theories based on GA are summarised in Table~\ref{tab:power_contributions}.
%
\vspace{-0.15cm}
\section{GAPoT Framework and Methodology}
\subsection{Circuit Analysis with GA}
\begin{figure} 
	\centering 
	\vspace{-0.3cm}
	\includegraphics[width=0.73\columnwidth]{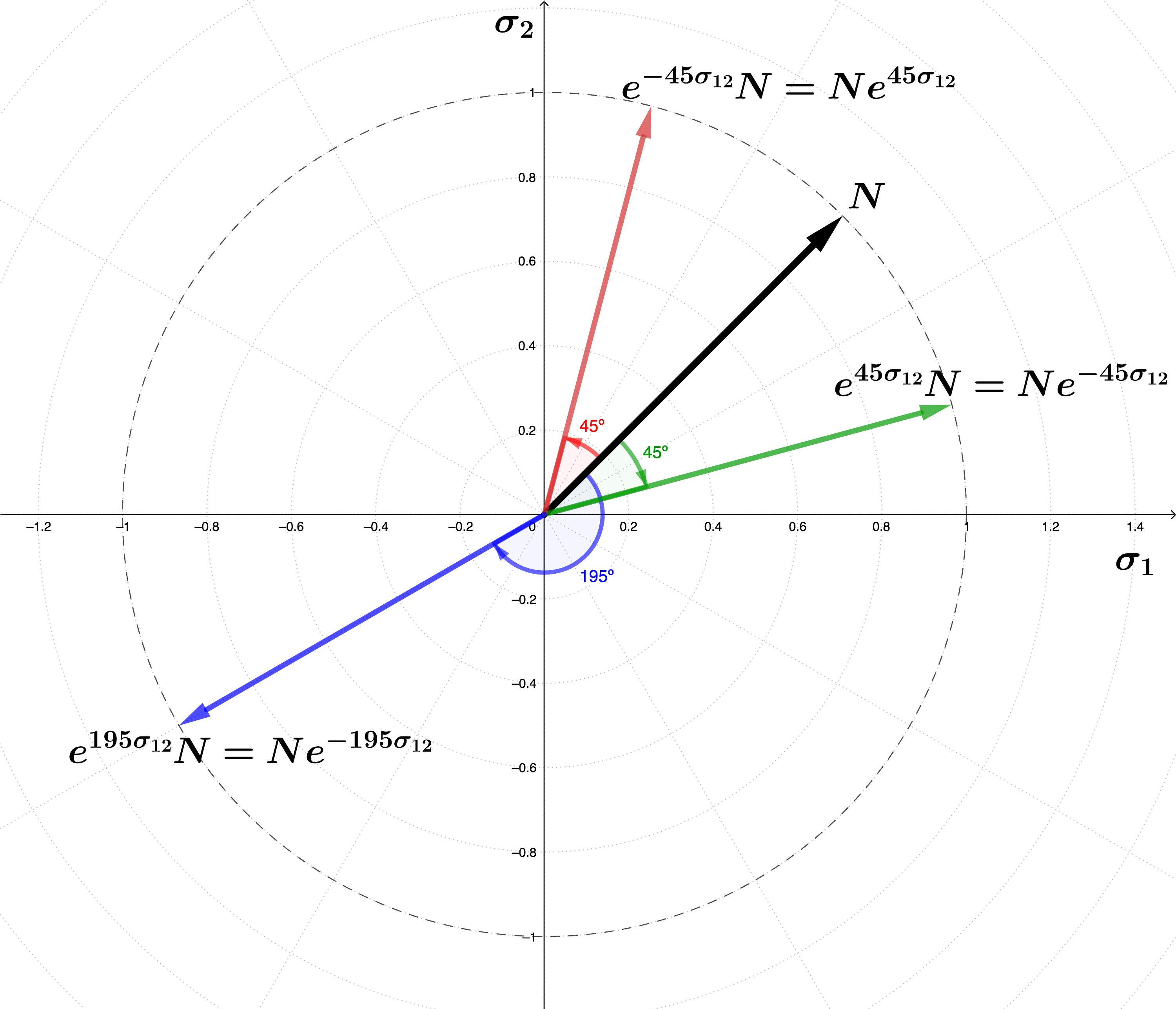}
	\vspace{-0.4cm}
	\caption{A vector multiplied by a rotor $(e^{\varphi\bm{\sigma}_{12}})$ rotates in clock- or counter-clock-wise direction depending on the type of multiplication.}
	\vspace{-0.45cm}
	\label{fig:rotor}
\end{figure}

%
%
In this theory, different approaches already available in the literature are unified and enhanced in order to analyse electrical circuits in the geometric domain.
%
%
The proposed modifications give full physical meaning to basic principles in electrical circuits.
%
%
An orthonormal basis is used in order to represent the multi-component nature of periodic signals with finite energy:
$\bm{\sigma}=\{\bm{\sigma}_{1},\bm{\sigma}_{2},\ldots,\bm{\sigma}_{n}\}$.
For example, for a voltage signal $u(t)$:
\vspace{-0.05cm}
\begin{equation}
\begin{aligned}
	u(t)&=U_0+\sqrt{2} { \textstyle \sum_{k=1}^{n}}U_k\sin(k \omega t + \varphi_k) \\ 
	&+ \sqrt{2} { \textstyle \sum_{l\in L}}U_l\sin(l \omega t + \varphi_l)
\end{aligned}
\label{eq:voltage_time_domain}
\end{equation}
where $U_0$ is the DC component, while $U_k$ and $\varphi_k$ are the RMS and phase of the $k$th harmonic, respectively.
%
The set $L$ represents possible sub- and inter-harmonics present in the signal~\cite{MONTOYA2019486}.
%
As in traditional circuit analysis based on complex variables, a rotating vector (similar to $e^{j\omega t}$) can be defined. This would facilitate later analyses in the geometric domain.
%
In addition, thanks to the linear properties of GA, it is possible to define a single multivector that includes all the harmonic frequencies present in the signal (this is not possible by using the traditional complex variable).
%
%
A rotating vector $\bm{n}(t)$ in a two-dimensional geometric space $\mathcal{G}_{2}$ can be obtained as follows~\cite{bernard1989multivectors}: 
%
\begin{equation}
\begin{aligned}
\bm{n}(t)&=e^{\frac{1}{2}\omega t\bm{\sigma}_{12}}\bm{N}e^{-\frac{1}{2}\omega t\bm{\sigma}_{12}}=\bm{R}\bm{N}\bm{R}^{\dagger} \\ &=e^{\omega t\bm{\sigma}_{12}}\bm{N}=\bm{R}^{2}\bm{N}=\bm{N}\bm{R}^{\dagger 2}
\end{aligned}
\label{eq:rotor_new}
\end{equation}
where $\bm{R}=e^{\frac{1}{2}\omega t\bm{\sigma}_{12}}$ is a~\textit{rotor} \cite{hitzer2013introduction} and $\bm{N}$ is a vector or~\textit{geometric phasor}. In (\ref{eq:rotor_new}), left-multiplying produces opposite effects compared to right-multiplying.
%
Fig.~\ref{fig:rotor} shows a graphical representation of a vector left-multiplied by a rotor (in green). This operation produces a rotation in clock wise direction. Similarly, a vector right-multiplied by a rotor (in red), produces a rotation in counter-clock wise direction.
%
%
%
In order to maintain the commonly accepted convention on signs in electrical engineering, vectors are left-multiplied by $e^{\omega t\bm{\sigma}_{12}}$. 
%
Therefore, a positive sign in an angle refers to the clock-wise direction.
This implies that inductors impedance will have positive angles while capacitors will have negative angles.
However, the  phase lead and lag changes its role: lag implies rotation in counter-clock wise direction and lead in clock-wise direction (see Fig.~\ref{fig:rotor}).
%
It can be readily demonstrated that the projection of a voltage $\bm{u}_1(t)$ over the basis $\bm{\sigma}_1$ yields the original voltage waveform, i.e., $u_1(t)=\sqrt{2}(\alpha_1\cos\omega t+\alpha_2\sin\omega t)$. This is the same as extracting the real part of the complex rotating vector, i.e. $\mathcal{R}e\{\sqrt{2}\vec{\bm{V}}e^{j\omega t}\}$. 
In fact, by using the orthonormal basis~$\bm{\sigma}=\{\bm{\sigma}_1 \rightarrow \sqrt{2}\cos\omega t;\bm{\sigma}_2 \rightarrow \sqrt{2}\sin\omega t \}$, $u_1(t)$ gets transformed into $\bm{u}_1=\alpha_1\bm{\sigma}_1+\alpha_2\bm{\sigma}_{2}$. Therefore:
\vspace{+0.1cm}
\begin{flalign}
& \bm{u}_1(t) =e^{\omega t \bm{\sigma}_{12}}\bm{u}_1=(\cos\omega t + \sin \omega t \bm{\sigma}_{12})(\alpha_1\bm{\sigma}_1+\alpha_2\bm{\sigma}_{2}) \nonumber \\
& =\left(\alpha_1\cos\omega t + \alpha_2\sin \omega t\right)\bm{\sigma}_1+\left(\alpha_2\cos\omega t - \alpha_1\sin \omega t\right)\bm{\sigma}_2 \nonumber
\\
& = \frac{1}{\sqrt{2}}\left(u_1(t)\bm{\sigma}_1-\bm{\bm{\bm{\mathcal{H}}}}\left[u_1(t)\right]\bm{\sigma}_2\right)
\label{eq:vector_giratorio}
\end{flalign}
where $\bm{\mathcal{H}}$ refers to the Hilbert transform of a signal~\cite{lev2006decomposition}.
%
Therefore, $u_1(t)={proj}_{\bm{\sigma}_1}[\sqrt{2}\bm{u}_1(t)]=\sqrt{2}\bm{u}_1(t)\cdot\bm{\sigma}_1$ is the projection of a rotating vector $\bm{u}_1(t)$ into $\bm{\sigma}_1$.
%
It is worth pointing out that the rotating vector $\bm{u}_1(t)$ is not the voltage itself, $u_1(t)$. This is a different interpretation compared to that of other authors~\cite{castro2012ieee,castro2016m}. This discrepancy will be analysed by using the circuit depicted in Fig.~\ref{fig:RLC_load}.
%
%

The time-domain equation that governs the circuit dynamics is:
\begin{equation}
u_1(t)
=Ri(t)+L\frac{di(t)}{dt}+\frac{1}{C}\int i(t)dt
\label{eq:RLC_circuit}
\end{equation}
%
Time derivatives in GA should be calculated as follows~\cite{bernard1989multivectors}:
%
\begin{equation*}
\begin{aligned}
\frac{d\bm{u}_1(t)}{dt}
&=
\omega \bm{\sigma}_{12}\bm{u}_1(t) \\
\int{\bm{u}_1(t)dt}
&=
-\frac{\bm{\sigma}_{12}}{\omega} \bm{u}_1(t)
\end{aligned}
\end{equation*}
If the source is sinusoidal and the circuit is operating in steady-state, (\ref{eq:vector_giratorio}) can be substituted in (\ref{eq:RLC_circuit}), yielding:
%
\begin{equation}
\begin{split}
proj_{\bm{\sigma}_1}
\left[ \sqrt{2}e^{\omega t \bm{\sigma}_{12}}\bm{u}_1 \right]
=
\sqrt{2}proj_{\bm{\sigma}_1}
\left[ Re^{\omega t \bm{\sigma}_{12}}\bm{i} \right.\\ 
\left.+ L\omega \bm{\sigma}_{12}e^{\omega t \bm{\sigma}_{12}}\bm{i}
- 
\frac
{\bm{\sigma}_{12}} 
{C\omega}e^{\omega t \bm{\sigma}_{12}}\bm{i}\right]
\end{split}
\label{eq:proyeccion}
\end{equation}
Equation (\ref{eq:proyeccion}) can be simplified, yielding
%
\begin{equation}
	\bm{u}_1
	=
	R\bm{i}+L\omega \bm{\sigma}_{12}\bm{i}
	-
	\frac
	{\bm{\sigma}_{12}}
	{C\omega}\bm{i}
	\label{eq:geom_RLC}
\end{equation}
Rotors $e^{\omega t \bm{\sigma}_{12}}$ are cancelled out because they commute with $\bm{\sigma}_{12}$. Therefore, it is not necessary to set any specific time instant, $t_0$, after performing the derivative, as suggested by CN.
%
The result is an algebraic equation where only geometric phasors such as $\bm{u}_1$ and $\bm{i}$ are present.
%
The {\em geometric impedance} can be obtained right-multiplying (\ref{eq:geom_RLC}) by the inverse of the current:
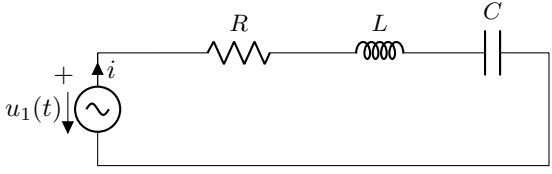
\begin{figure}[!t]
	\centering 
	\begin{circuitikz}[scale=1.5,/tikz/circuitikz/bipoles/length=1cm] \draw
		(0,0) to[sV, v<=$u_1(t)$, i=$i$] (0,1)  --  (0.5,1) to[R, l^=\small $R$] (2,1)		
		(2,1)  to [L, cute inductors,  l^=\small $L$] (3,1)
		(3,1)  to [C, cute inductors, l^=\small $C$] (4,1)-- (4,0) -- (0,0)
		;
		
		\draw (-0.3,0.8) node{$+$};
	\end{circuitikz}
	\vspace{-0.1cm}
	\caption{$RLC$ circuit used in Example 1.}
	\vspace{-0.4cm}
	\label{fig:RLC_load}
\end{figure}
%
%
\begin{equation}
	\bm{Z}
	=
	\bm{u}_1\bm{i}^{-1}
	=
	R+\left( L\omega-\frac{1}{C\omega} \right) \bm{\sigma}_{12}
	=
	R+X\bm{\sigma}_{12}	
\label{eq:z}
\end{equation}
The {\em geometric admittance} can be defined as the inverse of the geometric impedance:
%
\vspace{-0.1cm}
\begin{equation}
\bm{Y}=\bm{Z}^{-1}
=
\frac
{\bm{Z}^{\dagger}}
{\bm{Z}^{\dagger}\bm{Z}}
=
\frac
{\bm{Z}^{\dagger}}
{\|\bm{Z}\|^2}
=
G+B\bm{\sigma}_{12}
\label{eq:y}
\end{equation}
Both elements have similar definitions to those of impedance/admittance in the complex domain.
%
However, here both are multivectors because they consist of a scalar part plus a bivector.
%
The use of this criterion allows to overcome the drawbacks of other theories in which inductive reactance was negative while capacitive reactance was positive~\cite{castro2010use}.

In order to transform the signal (\ref{eq:voltage_time_domain}) from the time to the geometric domain, the following basis can be defined:
%
\vspace{-0.1cm}
\begin{equation}
\allowdisplaybreaks
\arraycolsep=1pt\def\arraystretch{1.1}
\begin{array}{lcl}
\varphi_{dc} = 1 \quad &\longleftrightarrow  & \quad  \bm{\sigma}_0\\
\varphi_{c1}(t) = \sqrt{2}\cos{\omega t} \quad &\longleftrightarrow  & \quad  \bm{\sigma}_1 \\
\varphi_{s1}(t) = \sqrt{2}\sin{\omega t} \quad & \longleftrightarrow & \quad  \bm{\sigma}_2 \\
& \vdots\\
\varphi_{cn}(t) = \sqrt{2}\cos n\omega t \quad & \longleftrightarrow & \quad  \boldsymbol{\sigma}_{2n-1} \\
\varphi_{sn}(t) = \sqrt{2}\sin n\omega t \quad & \longleftrightarrow & \quad  \boldsymbol{\sigma}_{2n} \\
\end{array}
\label{eq:GAPOT_transformation}
\end{equation}
\noindent In addition, $l$ sub- and inter-harmonics can be added by increasing the number of elements in the basis by $2l$ after the highest-order harmonic $(n)$~\cite{MONTOYA2019486}.
%
%
The voltage $u(t)$ in (\ref{eq:voltage_time_domain}) can be transformed to the geometric domain:
\begin{equation}
\begin{aligned}
	\bm{u}&=U_0\bm{\sigma}_0+
	{\textstyle \sum_{k=1}^{n}}U_ke^{-\varphi_k\bm{\sigma}_{(2k-1)(2k)}}\bm{\sigma}_{(2k-1)} \\ &+
	{\textstyle  \sum_{m=1}^{l}}U_me^{-\varphi_m\bm{\sigma}_{(2n+2m-1)(2n+2m)}}\bm{\sigma}_{(2n+2m-1)}\\
	&=U_0+{\textstyle \sum_{k=1}^{n}}U_{k1}\bm{\sigma}_{(2k-1)}+U_{k2}\bm{\sigma}_{(2k)}\\
	&+{\textstyle \sum_{m=1}^{l}}U_{m1}\bm{\sigma}_{(2m-1)}+U_{m2}\bm{\sigma}_{(2m)}
\end{aligned}
\label{eq:u_geometric}
\end{equation}
\noindent where $U_{k1}=U_k\cos\varphi_k$ and $U_{k2}=U_k\sin\varphi_k$. 
%
The same transformation can be applied to $i(t)$ in order to calculate $\bm{i}$.
It is worth noting that $\bm{i}$ may include harmonics not present in the voltage.
%
%
By using the same rationale presented in (\ref{eq:RLC_circuit})-(\ref{eq:geom_RLC}), the geometric impedance can be defined for each harmonic as:
%
\begin{equation}
\bm{Z}_k
=
\bm{u}_k\bm{i}_k^{-1}
=
R
+ 
\left( kL\omega-\frac{1}{kC\omega}\right)
\bm{\sigma}_{(2k-1)(2k)}	
\label{eq:general_z}
\end{equation}
\noindent where $\bm{u}_k$ and $\bm{i}_k$ are geometric phasors for the harmonic $k$.

This proposal overcomes some drawbacks of previous GA power theories. First, it can readily accommodate DC components in voltages and currents. Second, the traditional definition of apparent power based on the product of the RMS voltage and current is preserved, and this does not happens in other proposals~\cite{castro2012ieee}. These are contributions of this work.
%
\vspace{-0.1cm}
\subsection{Power Flow in GA}
There exist different definitions for apparent power in power theories based on GA.
%
Menti and Castro-N\'{u}ñez chose $\bm{S}=\bm{UI}$, while Castilla-Bravo chose $\bm{S}=\bm{UI}^{*}$.
%
All of them are compatible with the energy conservation principle due to the multi-component nature of GA~\cite{castro2012advantages}.
%
However, results might be inconsistent if the orthonormal basis that expands the geometric space is not carefully chosen.
%
For example, in the proposal of CN, $k$-vectors are used for the basis~\cite{castro2010use}. Therefore, the geometric power calculation should be adapted so that power components are correctly computed, as already mentioned in Section~\ref{sec:overview}.
%
%
%
Also, non-active power calculations can lead to erroneous results since the geometric power is not calculated as $\bm{M}=\bm{U}\bm{I}^{\dagger}$~\cite{montoya2020geometric}.
%
%
%
In order to prove it, the apparent geometric power can be defined as:
%
\begin{equation}
\bm{M}
=
\bm{u}\bm{i}
=
\bm{u}\cdot\bm{i}
+
\bm{u}\wedge\bm{i} 
\label{eq:M_power}
\end{equation}
The value of $\|\bm{M}\|$ is the product of the voltage and current modules, provided that $\bm{u}$ and $\bm{i}$ are vectors:
%
\begin{equation}
\begin{aligned}
	\|\bm{M}\|
	&=
	\sqrt{\langle\bm{M}^{\dagger}\bm{M}\rangle_0}
	=
	\sqrt{\langle\left(\bm{ui}\right)^{\dagger}\left(\bm{ui}\right)\rangle_0} \\
	&=
	\sqrt{\langle\left(\bm{i}^{\dagger}\bm{u}^{\dagger}\right)\left(\bm{ui}\right)\rangle_0}
	=
	\sqrt{\|\bm{u}\|^2\|\bm{i}\|^2}
	=
	\|\bm{u}\|\|\bm{i}\|
	\end{aligned}
\end{equation}
\noindent where the property $\bm{a}^{\dagger}=\bm{a}$ has been applied for vectors.
The application of this property is the key to overcome a definition based on the complex conjugate current.
%
%
This feature cannot be applied in other power theories based on GA since, in general, $\bm{A}^{\dagger }\neq \bm{A}$ for any $k$-vector $\bm{A}$ with $k>1$~\cite{castro2010use}.

In (\ref{eq:M_power}), several terms of engineering interest can be identified.
%
On the one hand, the scalar term $\langle\bm{M}\rangle_0=\bm{u}\cdot\bm{i}$ matches the active power $P$, and it will be called active geometric power, or $M_a$. 
%
On the other hand, the bivector term $\langle\bm{M}\rangle_2=\bm{u}\wedge\bm{i}$ will be called non-active geometric power, or $\bm{M}_N$. 






%
\vspace{-0.2cm}
\subsection{Current Decomposition in GA}
In this section, the current consumed by a load is decomposed by using the proposed power theory.
Simplifying (\ref{eq:M_power}) and taking into account that for any vector $\bm{a}^{-1}={\bm{a}}/{\|\bm{a}\|^2}$:
%
\begin{equation}
\begin{aligned}
	&\bm{M}=\bm{ui} \quad \longrightarrow \quad \bm{u}^{-1}\bm{M}=\underbrace{\bm{u}^{-1}\bm{u}}_{1}\bm{i}=\bm{i} \\ &\bm{i}=\bm{u}^{-1}\bm{M}=\frac{\bm{u}}{\|\bm{u}\|^2}\left(M_a+\bm{M}_N\right)=\bm{i}_a+\bm{i}_N
\end{aligned}
	\label{eq:ia_iN}
\end{equation}
\noindent where $\bm{i}_a$ is the active or Fryze current, while $\bm{i}_N$ is the non-active current. 
%
This decomposition procedure has not been used before for GA power theories in the frequency domain, and it is a novel contribution of this work.
%
Also, in previous power theories based on GA current decomposition was not guaranteed since multivectors might not have inverse and, in any case, its calculation is not straightforward~\cite{hitzer2019construction}.

Each of the currents presented above has a well-established engineering meaning.
%
The current $\bm{i}_a$ is the minimum current required to produce the same active power to that consumed by the load, while the non-active current $\bm{i}_N$ is the current that does not affect the net active power. Therefore, the latter can be compensated by using either passive or active filters.
%
The current $\bm{i}_N$ can be decomposed in two terms for practical engineering purposes. 
%
The first one is related to transient energy storage and leads to the reactive current. The second one does not include storage and leads to the scattered current introduced by Czarnecki~\cite{czarnecki2009compensation}.
%
In addition, by using (\ref{eq:y}), (\ref{eq:u_geometric}) and Ohm's law, the current $\bm{i}$ demanded by a linear load can be calculated as:
%
\begin{equation}
	\bm{i}=\sum_{k=1}^{n}\bm{Y}_k\bm{u}_k=\sum_{k=1}^{n}\left(G_k+B_k\bm{\sigma}_{(2k-1)(2k)}\right)\bm{u}_k=\bm{i}_p+\bm{i}_q
	\label{eq:ip_iq}
	\nonumber
\end{equation}
\noindent where $\bm{i}_p$ is commonly known as {\em parallel current} while $\bm{i}_q$ as {\em quadrature current}:
%
\begin{equation}
	\begin{aligned}
	\bm{i}_p
	=
	\sum_{k=1}^{n}G_k\bm{u}_k,\;\;
	\bm{i}_q
	=
	\sum_{k=1}^{n}B_k\bm{\sigma}_{(2k-1)(2k)}\bm{u}_k
	\end{aligned}
\end{equation}
It can be readily demonstrated that they are orthogonal. 
Therefore, by comparing (\ref{eq:ia_iN}) and (\ref{eq:ip_iq}):
%
\begin{equation}
	\bm{i}=\bm{i}_a+\bm{i}_N=\bm{i}_p+\bm{i}_q=\bm{i}_a+\bm{i}_s+\bm{i}_q
	\label{eq:decomposition_currents}
\end{equation}
\noindent where $\bm{i}_s=\bm{i}_p-\bm{i}_a$ is the scattered current,
%
which can only be compensated by using active elements, while $\bm{i}_q$ can be compensated by using passive elements~\cite{czarnecki2009compensation}.
%
There have been different attempts to give physical meaning to these current components. For that purpose, the scattered power was defined as $\bm{M}_s=\bm{u}\bm{i}_s$, while reactive power as $\bm{M}_q=\bm{u}\bm{i}_q$.
%
However, it has already been demonstrated that this decomposition has no physical meaning, even though is useful for the engineering practice~\cite{cohen1999physical,de2010ac}.
%
In addition, the component $\bm{i}_G$ is included to model current components whose frequencies are not present in the voltage:
%
\begin{equation}
\bm{i}
=
\bm{i}_a+
\underbrace{\bm{i}_s+
\bm{i}_q+
\bm{i}_G}_{\bm{i}_N}
\label{eq:decomposition_currents_generated}
\end{equation}
The power factor can be defined in the geometric domain as:
%
\begin{equation}
	pf=\frac{\langle\bm{M}\rangle_0}{\|\bm{M}\|}=\frac{P}{\|\bm{M}\|}
\end{equation}
\section{Examples and Discussion}
Two examples will be given in order to validate the theoretical developments. The first one is the resolution of an $RLC$ circuit under distorted conditions, while the other one consist on the analysis of experimental data.
%
The results obtained with the proposed theory will be compared to those obtained by other theories.
%
All the results have been obtained by using Matlab and the Clifford Algebra toolbox~\cite{sangwine2017clifford}.
%
\vspace{-0.1cm}
\subsection{Example 1. Non-Sinusoidal Source}
The $RLC$ circuit presented in Fig.~\ref{fig:RLC_load} has been used as an example and benchmark for the different theories based on GA.
%
First, Menti, Castilla-Bravo and Lev-Ari theories cannot address it since they do not offer the tools for analysing circuits in the geometric domain.
%
For these cases, it would be required to solve the circuit by using other techniques (such as complex algebra), and then transform the results to the geometric domain in order to analyse the power flow.
%
Therefore, the circuit will only be solved by using the theory proposed in this paper (GAPoT), CN~\cite{castro2010use} and CPC (Czarnecki)~\cite{czarnecki2005currents}. All of them allow to decompose the current.
%

In the circuit, $R=1\;\Omega$, $L=1/2$~H and $C=2/3$~F. The source voltage is $u(t)=100\sqrt{2}\left(\sin \omega t + \sin 3\omega t\right)$. 
Kirchhoff laws can be applied in the time domain in order to obtain Equation (\ref{eq:RLC_circuit}).
%
Then, GAPoT theory is used to transform it to the geometric domain:
\begin{equation*}
\begin{aligned}
	\bm{u}_1 +\bm{u}_3
	&=
	R\left(\bm{i}_1+\bm{i}_3\right)
	+L\left(\omega \bm{\sigma}_{12}\bm{i}_1+3\omega  \bm{\sigma}_{56}\bm{i}_3\right) \\
	&-\frac{1}{C}\left(\frac{\bm{\sigma}_{12}\bm{i}_1}{\omega }+\frac{\bm{\sigma}_{56}\bm{i}_3}{3\omega }\right)
\end{aligned}
\end{equation*}
It can be seen that the superposition theorem is embedded in the proposed formulation since all components are operated at the same time. This is a clear difference compared to theories based on complex numbers.

%
%
By using (\ref{eq:GAPOT_transformation}), the geometric voltage turns into:
\begin{equation}
\bm{u}
=
\bm{u}_1+\bm{u}_3
=
100\left(\bm{\sigma}_2+\bm{\sigma}_6\right)
\label{ec.def.valor.tension}	
\end{equation}
\noindent while impedances and admittances are calculated with (\ref{eq:general_z}):
%
\begin{equation*}
\begin{aligned}
	\bm{Z}_1&=1-\bm{\sigma}_{12} \longrightarrow \bm{Y}_1=0.5+0.5\bm{\sigma}_{12} \\
	\bm{Z}_3&=1+\bm{\sigma}_{56} \longrightarrow \bm{Y}_3=0.5-0.5\bm{\sigma}_{56}
\end{aligned}
\end{equation*}
Therefore, the current becomes:
%
\begin{equation*}
\begin{aligned}
	\bm{i}&=\bm{i}_1+\bm{i}_3=\bm{Y}_1\bm{u}_1+\bm{Y}_3\bm{u}_3 \\
	&=50\bm{\sigma}_1+50\bm{\sigma}_2-50\bm{\sigma}_5+50\bm{\sigma}_6
	\end{aligned}
\end{equation*}
The geometric apparent power is calculated by using (\ref{eq:M_power}):
%
\begin{equation}
	\bm{M}=\bm{ui}=\underbrace{10}_{M_a=P}-\underbrace{5\bm{\sigma}_{12}+5\bm{\sigma}_{56} - 5\bm{\sigma}_{16}-5\bm{\sigma}_{25}}_{\bm{M}_N}\;\text{kVA}
	\nonumber
\end{equation}
The active power consumption is 10~kW, while the rest is non-active power.
%
The reactive power consumed by each harmonic is included in $\bm{\sigma}_{(2k-1)(2k)}$.
%
Therefore, the reactive power of the first harmonic is $-5\bm{\sigma}_{12}$, while that of the third one is $5\bm{\sigma}_{56}$.
%
This result is in good agreement with traditional analyses in the frequency domain where the reactive power of each harmonic is the same, but with opposite sign.
However, the term $-5\bm{\sigma}_{16}-5\bm{\sigma}_{25}$ cannot be obtained by using complex algebra since it involves the cross-product between voltages and currents of different frequencies.
%
This is one of the advantages of GA.

The module of the geometric power is:
\begin{equation*}
	\|\bm{M}\|=\sqrt{\langle\bm{M}^{\dagger}\bm{M}\rangle_0}=\|\bm{u}\|\|\bm{i}\|=141.42\mathbin{\times} 100 = 14,142 \, \text{VA}
\end{equation*} 

If the CN theory is applied, the apparent power becomes:
%
\begin{equation}
	\begin{aligned}
		\bm{M}_{CN}=10 + 10\bm{\sigma}_{12}+10\bm{\sigma}_{34} \;\;\text{kVA}
	\end{aligned}
\end{equation}

The value of active power is 10~kW. However, the factor $f=(-1)^{k(k-1)/2}$ has been used for the calculations.
%
Also, it can be seen that it is not possible to distinguish reactive power components generated by each harmonic since all of them are grouped in the term $\bm{\sigma}_{12}$.
%
Finally, it can be observed that $\|\bm{M}_{CN}\|\neq \|\bm{u}\|\|\bm{i}\|$.

By using the CPC theory, it is not possible to generate a current vector in the frequency domain nor a power multivector. 
%
%
Also, the instantaneous value of currents should be used to describe independent terms of power. The results are:
%
%
\begin{equation*}
\begin{split}
&P = 10.000\;\text{W}
\;\;\; 
Q_r = 10.000\;\text{VAr}
\;\; \\
&D_s = 0\;\text{VA} 
\;\;\;\;\;\;\;\;
S = 14.142\;\text{VA}
\end{split}
\end{equation*}
The value of active power calculated by the CPC theory is, of course, correct.
However, this theory cannot fully describe harmonic interactions between voltage and current components.
%
The module of the total reactive power yields 10~kVAr. However, it is not possible to calculate the individual contribution of each harmonic nor its sign (sense).
%

Regarding current decomposition, by using (\ref{eq:ia_iN}), it follows:
%
\begin{equation*}
\bm{i}
=
\underbrace{50\bm{\sigma}_2
+
50\bm{\sigma}_6}_{\bm{i}_a}
+
\underbrace{50\bm{\sigma}_1
-
50\bm{\sigma}_5}_{\bm{i}_N}
\end{equation*}
Also, if (\ref{eq:ip_iq}) is applied, an identical result is obtained:
%
\begin{equation*}
\bm{i}
=
\underbrace{50\bm{\sigma}_2+50\bm{\sigma}_6}_{\bm{i}_p}
+
\underbrace{50\bm{\sigma}_1
-
50\bm{\sigma}_5}_{\bm{i}_q}
\end{equation*}
If a harmonic compensator is to be designed, its susceptance at each harmonic would be the same as that of the load, but with the opposite sign:
\begin{equation*}
B_{cp1}=-B_{1}
\quad
B_{cp3}=-B_{3}
\end{equation*}
All the current will be compensated by using passive elements since no scattered current is present (see~\cite{montoya2019quadrature} for more details). This means that $\bm{i}_a=\bm{i}_p$. Therefore, $\bm{i}_N$ would be zero. 
%
%
%
%

Consider now a value of $C=2/7$~F in Fig.~\ref{fig:RLC_load}.
%
This set of parameters has been used in other scientific works since power components cannot be distinguished if the classical concept of apparent power is applied~\cite{czarnecki1997budeanu,castro2012advantages}.
%
%
For the voltage value presented in (\ref{ec.def.valor.tension}), the current becomes:
%
\vspace{-0.1cm}
\begin{equation*}
\bm{i}=30\bm{\sigma}_1+10\bm{\sigma}_2-30\bm{\sigma}_5+90\bm{\sigma}_6
\end{equation*}
\vspace{-0.1cm}
\noindent and the geometric power is:
\vspace{-0.1cm}
\begin{equation*}
	\bm{M}=10-3\bm{\sigma}_{12}+3\bm{\sigma}_{56}-3\bm{\sigma}_{16}-3\bm{\sigma}_{25}+8\bm{\sigma}_{26}\;\;\text{kVA}
\end{equation*}
\vspace{-0.05cm}
Active power consumption is the same to that obtained with other theories (10~kW). However, the rest of terms are different.
%
Reactive power consumption for each harmonic has been reduced. 
%
The term $8\bm{\sigma}_{26}$ has appeared due to the interaction between in-phase components in the first voltage harmonic and the third current harmonic.
%
This term highlights that the system cannot be fully compensated by using only passive elements.
%
Despite the changes in various terms in current and powers, the module of the geometric power remains unchanged:
%
\begin{equation*}
\|\bm{M}\|=\|\bm{u}\|\|\bm{i}\|=141.42\mathbin{\times} 100 = 14.14 \, \;\text{kVA}
\end{equation*} 
The current decomposition for this case is given in Table~\ref{tab:currents_non_sinusoidal}.
%

If the CN theory is applied, the power becomes:
\begin{table}
	\centering
	\begin{tabular}{@{}lrrrrrrr@{}}
		\toprule
		& \multicolumn{1}{c}{$\bm{\sigma_1}$} & \multicolumn{1}{c}{$\bm{\sigma_2}$} & \multicolumn{1}{c}{$\bm{\sigma_{3}}$} & \multicolumn{1}{c}{$\bm{\sigma_{4}}$} & \multicolumn{1}{c}{$\bm{\sigma_{5}}$} & \multicolumn{1}{c}{$\bm{\sigma_{6}}$} & \multicolumn{1}{r}{$\|\cdot\|$} \\ \cmidrule(l){2-7} \cmidrule(l){8-8}
		$\bm{i_a}$ & 0       & 50.00    & 0   & 0    & 0  & 50.00  & 70.71  \\
		$\bm{i_s}$ & 0       & -40.00    & 0   & 0    & 0  & 40.00  & 56.56  \\ \cmidrule{2-8}
		$\bm{i_p}$ & 0       & 10.00    & 0   & 0    & 0  & 90.00  & 90.55  \\
		$\bm{i_q}$ & 30.00       & 0    & 0   & 0    & -30.00  & 0  & 42.42  \\ \cmidrule{2-8}
		$\bm{i}$ & \textbf{30.00}   & \textbf{10.00}   & \textbf{0}   & \textbf{0}& \textbf{-30.00}   & \textbf{90.00}   & \textbf{100.00}\\ \bottomrule
	\end{tabular}
	\vspace{+0.1cm}
	\caption{Current decomposition for the circuit in Fig.~2 and C=2/7 F.}
	\vspace{-0.5cm}
	\label{tab:currents_non_sinusoidal}
\end{table}
%
%
\begin{equation}
\bm{M}=10+6\bm{\sigma}_{12}+6\bm{\sigma}_{34}+8\bm{\sigma}_{1234}\;\;\text{kVA}
\end{equation}
where $\|\bm{M}\|=15.36$~kVA. This value differs from that obtained in the previous case, even though voltages and currents have not changed.
%
%
Therefore, the proposed theory captures effects that others cannot (e.g. CN theory).
%
\vspace{-0.1cm}
\subsection{Example 2. Measurements Analysis}
\begin{figure}[!t]
	\centering 
	\includegraphics[width=0.93\columnwidth]{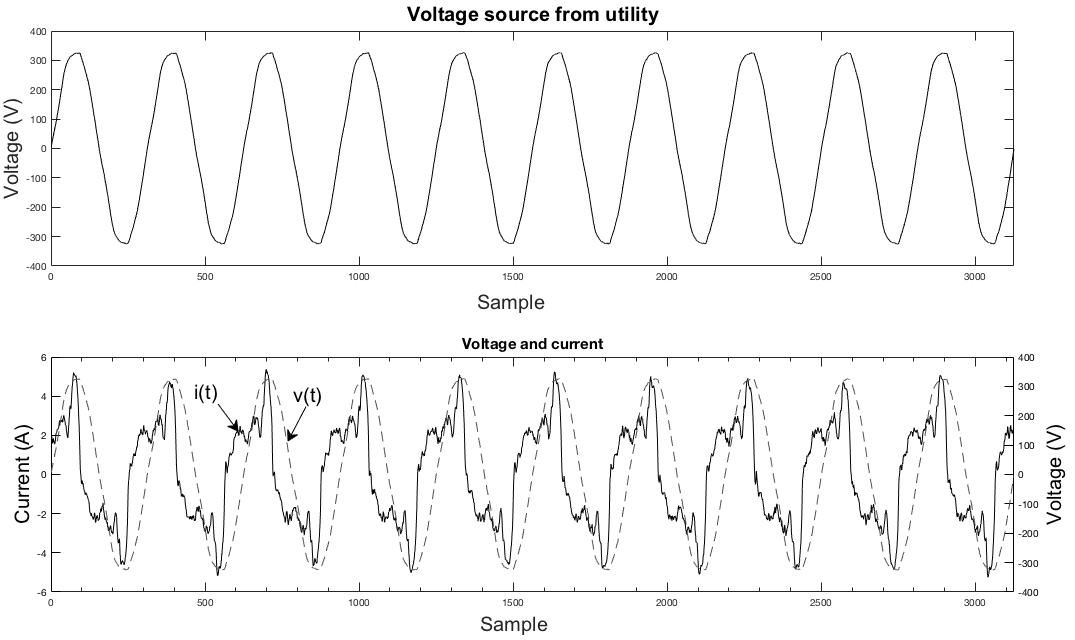}
	\vspace{-0.15cm}
	\caption{Voltage and current measurements from a house in Almer\'{\i}a, Spain.}
	\label{fig:ozm_voltage}
	\vspace{-0.15cm}
\end{figure}

In this example, the voltage and current waveforms of a typical house in Almer\'{\i}a (Spain) are analysed. The open-platform openZmeter~\cite{viciana2018openzmeter,viciana2019open} was used for analysing power quality.
Fig.~\ref{fig:ozm_voltage} shows voltage and current measurements in a time window of 200~ms, taken with a sampling frequency of 15.625~kHz (3125 samples). 
%
Several home appliances were on, like a TV, LED lights and electronic appliances  such as a router, satellite receiver and other devices in stand by mode.
%
The current waveform was highly distorted since the current THD was 88.3\%, while the voltage THD was 6.63\%.
%
\begin{figure}[!t]
	\centering 
	\vspace{-0.2cm}
	\includegraphics[width=0.93\columnwidth]{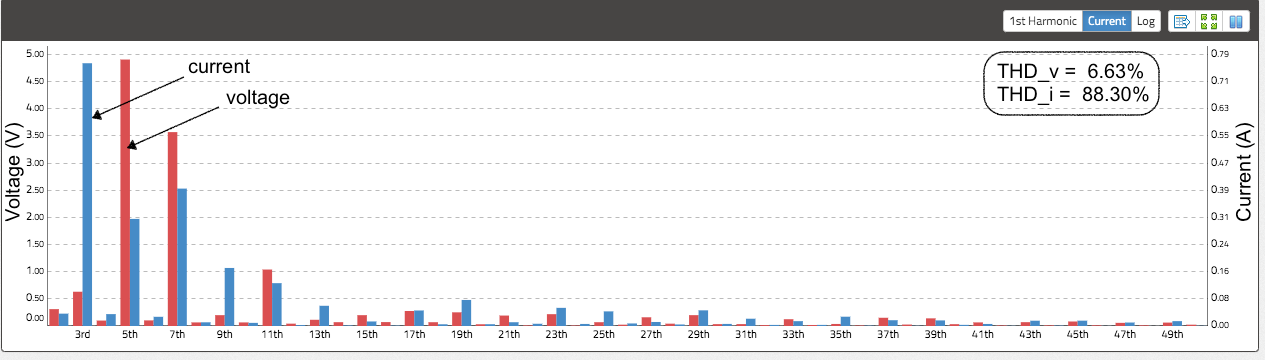}
	\vspace{-0.15cm}
	\caption{Voltage and current spectrum from a house in Almer\'{\i}a, Spain.}
	\vspace{-0.2cm}
	\label{fig:ozm_freq}
\end{figure}
\begin{figure}[!t]
\centering 
\vspace{-0.25cm}
\includegraphics[width=0.93\columnwidth]{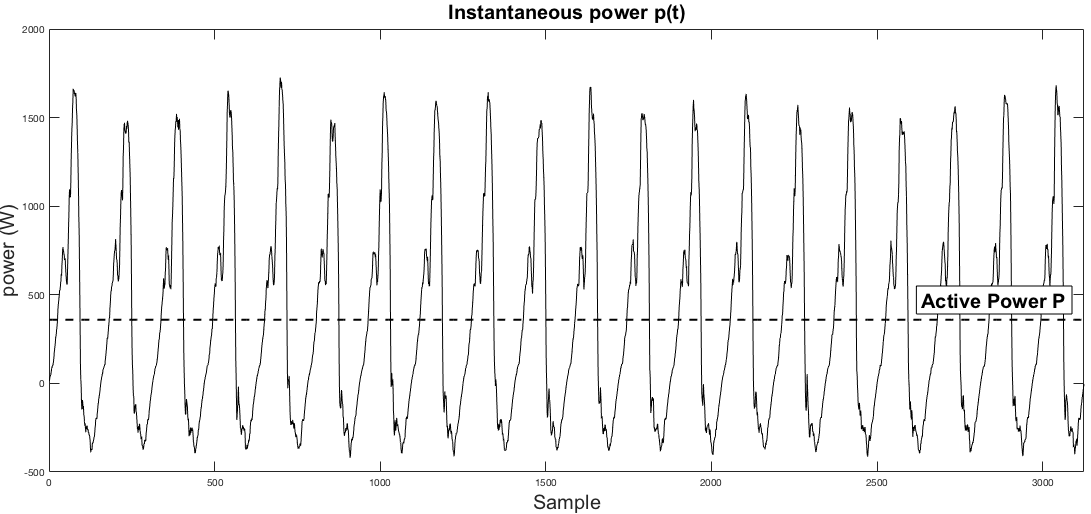}
\vspace{-0.15cm}
\caption{Instantaneous power waveform and active power $P$ in Example 2.}
\vspace{-0.1cm}
\label{fig:ozm_power}
\end{figure}
Fig.~\ref{fig:ozm_freq} shows the voltage and current spectrum for the first fifty harmonics (for the sake of clarity, the fundamental component is not shown).
%
%
Fifth and seventh harmonic voltage components are prominent while even harmonics are insignificant due to the half-wave symmetry of the waveform.
%
%
From Table~\ref{tab:harm}, it can be concluded that most of the energy is concentrated in the first five odd harmonics.
%
%
The RMS value of the voltage is 234.00V while that of the current is 2.61A.
%
Fig.~\ref{fig:ozm_power} shows the power waveform as well as the value of $P$ that is 359.15~W.

%

A geometric vector can be derived by using the data of Table~\ref{tab:harm}.
%
A value of $n=5$ has been considered (the fundamental component plus four harmonics). Therefore, the dimension of the geometric space has been set 10 ($2n$).
%
It is worth pointing out that in the proposed theory the dimension of the geometric space can be chosen according to the requirements (numbers of harmonics). This is an advantage compared to other theories.
%
%

\begin{figure}[!t]
\centering 
\includegraphics[width=0.42\textwidth]{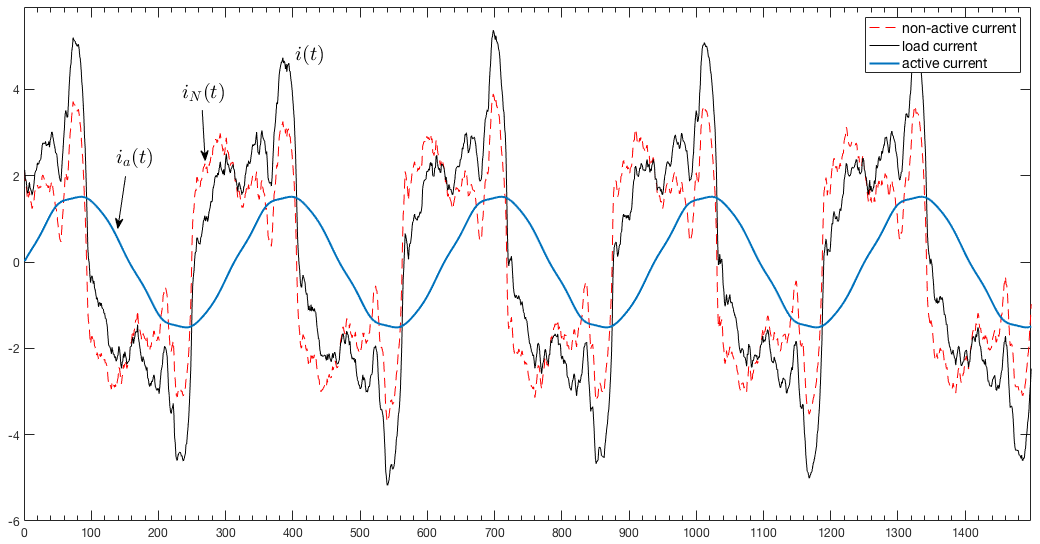}
\vspace{-0.1cm}
\caption{Total, active and non-active current for the measurements.}
\vspace{-0.3cm}
\label{fig:ozm_house_currents}
\end{figure}

The voltage and current expressions in polar form are:
\begin{equation*}
\begin{aligned}
\bm{u}&=233.92e^{-1.57\bm{\sigma}_{12}}\bm{\sigma}_1+0.46e^{-2.61\bm{\sigma}_{34}}\bm{\sigma}_3+4.74e^{1.28\bm{\sigma}_{56}}\bm{\sigma}_5 \\
&+4.02e^{-0.07\bm{\sigma}_{78}}\bm{\sigma}_7+0.42e^{-2.60\bm{\sigma}_{(9)(10)}}\bm{\sigma}_9 \\[0.2cm]
\bm{i}&=2.33e^{-0.72\bm{\sigma}_{12}}\bm{\sigma}_1+0.93e^{1.85\bm{\sigma}_{34}}\bm{\sigma}_3+0.45e^{-1.69\bm{\sigma}_{56}}\bm{\sigma}_5 \\
&+0.49e^{1.70\bm{\sigma}_{78}}\bm{\sigma}_7+0.16e^{-1.44\bm{\sigma}_{(9)(10)}}\bm{\sigma}_9
\end{aligned}
\end{equation*}
\noindent and the geometric power is:
%
\begin{equation}
\begin{aligned}
\bm{M}&=359.21+408.50\bm{\sigma}_{12}-0.42\bm{\sigma}_{34}-0.34\bm{\sigma}_{56} \\
&+1.95\bm{\sigma}_{78}+0.06\bm{\sigma}_{(9)(10)}+\bm{O}
\end{aligned}
\end{equation}
%
\noindent where $\bm{O}$ includes the rest of bivectors that appears due to the cross products.
The value of $M_a$ is $359.21$~W, which is similar to that obtained by using the digital samples of voltages and currents.
%
Results for the reactive power of each harmonic are also similar.
%
These values are shown in Table~\ref{tab:comparative_power_harm}. Table~\ref{tab:currents_real_circuit} shows the current components presented in (\ref{eq:ia_iN}).
%
%
In order to compute $\bm{i}_p$ and $\bm{i}_q$ according to (\ref{eq:general_z}), the geometric impedances were calculated for each harmonic.
%
%
The value of the total current was $\|\bm{i}\|=2.607$~A, while $\bm{i}_a=1.535$~A. The latter is the minimum current that would produce the same active power.
%
%
%
%
Fig.~\ref{fig:ozm_house_currents} shows the waveforms of $i$, $i_a$ and $i_N$.
\begin{table}[!t]
	\centering
\vspace{-0.1cm}
\begin{tabular}{lrrrr} 
		\toprule
      & \multicolumn{2}{c}{\textbf{Voltage}} & \multicolumn{2}{c}{\textbf{Current}} \\ \cmidrule(lr){2-3}\cmidrule(lr){4-5}
\textbf{Order} & $\|V\|$     (V)      & $\varphi_v$    (rad)   & $\|I\|$    (A)      & $\varphi_i$   (rad)     \\
      		\midrule
fund  & 233.92      & -1.57      & 2.33        & -0.72       \\
3rd   & 0.46         & -2.61      & 0.93        & 1.85        \\
5th   & 4.74         & 1.28       & 0.45        & -1.69       \\
7th   & 4.02         & -0.07      & 0.49        & 1.70        \\
9th   & 0.42         & -2.60      & 0.16        & -1.44 \\
		\bottomrule
\end{tabular}
\vspace{+0.1cm}
\caption{Odd harmonics present in the waveforms of Example 2.}
\vspace{-0.4cm}
\label{tab:harm}
\end{table}

\begin{table}[!t]
	\centering
	\begin{tabular}{lrrr} 
		\toprule

      & \multicolumn{1}{c}{$\bm{P}_i$}      & \multicolumn{2}{c}{$\bm{Q}_i$} \\ \cmidrule(l){2-2} \cmidrule(lr){3-4}
\textbf{Order} & \multicolumn{1}{c}{\textbf{ozm}}           & \multicolumn{1}{c}{\textbf{ozm}}              & \multicolumn{1}{c}{\textbf{GA}}       \\
		\midrule
fund  & 361.80  & 408.56   & 408.50   \\
3rd   & -0.102 & -0.426   & -0.425    \\
5th   & -2.134 & -0.346   & -0.346    \\
7th   & -0.408 & 1.955    & 1.955     \\
9th   & 0.028  & 0.063    & 0.062     \\
\midrule
\textbf{Total} & \textbf{359.15}       &                &     \\
		\bottomrule
\end{tabular}
\vspace{+0.1cm}
\caption{Harmonic active (W) and reactive (VAr) power measurements.}
\vspace{-0.4cm}
\label{tab:comparative_power_harm}
\end{table}

\begin{table}[!t]
	\centering
	\begin{tabular}{@{}lrrrrrr@{}}
		\toprule
		& \multicolumn{1}{c}{$\bm{i}_p$} & \multicolumn{1}{c}{$\bm{i}_a$} & \multicolumn{1}{c}{$\bm{i}_s$} & \multicolumn{1}{c}{$\bm{i}_q$}& \multicolumn{1}{c}{$\bm{i}_N$} & \multicolumn{1}{c}{$\bm{i}$} \\ \cmidrule(l){2-7} 
		$\bm{\sigma}_1$ &   -0.007     & -0.007    & 0.000   & 1.746    & 1.746  & \textbf{1.739}    \\
		$\bm{\sigma}_2$ & 1.547       & 1.534    & 0.012   & 0.008    & 0.020  & \textbf{1.555}    \\
		$\bm{\sigma}_3$ & 0.188       & -0.003    & 0.190   & -0.454    & -0.263  & \textbf{-0.266}    \\
		$\bm{\sigma}_4$ & -0.108       & 0.001    & -0.109   & -0.789    & -0.898  & \textbf{-0.897}    \\
		$\bm{\sigma}_5$ & -0.126       & 0.009    & -0.135   & 0.070    & -0.065  & \textbf{-0.056}    \\
		$\bm{\sigma}_6$ & 0.431       & -0.030    & 0.461   & 0.020    & 0.482  & \textbf{0.452}    \\
		$\bm{\sigma}_7$ & -0.101       & 0.026    & -0.127   & 0.036    & -0.091  & \textbf{-0.065}    \\
		$\bm{\sigma}_8$ & -0.007       & 0.002    & -0.010   & -0.484    & -0.494  & \textbf{-0.492}    \\
		$\bm{\sigma}_9$ & -0.057       & -0.002    & -0.055   & 0.077    & 0.022  & \textbf{0.020}    \\
		$\bm{\sigma}_{10}$ & 0.034       & 0.001    & 0.033   & 0.129    & 0.162  & \textbf{0.163}    \\ \cmidrule{2-7}
		$\|\cdot\|$ & \textbf{1.629}       & \textbf{1.535}     & \textbf{0.548}    & \textbf{2.035}     & \textbf{2.108}   & \textbf{2.607}     \\ \bottomrule
	\end{tabular}
\vspace{+0.1cm}
\caption{Current components obtained from current measurements.}
\vspace{-0.6cm}
\label{tab:currents_real_circuit}
\end{table}
%
\vspace{-0.15cm}
\section{Extension to Multi-Phase Systems}
It might be possible to extend the proposed theory to multi-phase systems thanks to the use of $N$-dimensional vectors.
%
In GA, the different phases can be operated as vector arrays in the geometric space.
%
For example, for a three-phase system, arrays of dimension three can be used, while each element of the array would be a vector of dimension $N$, depending on the number of harmonics to be considered.
%
In particular, voltages and currents could be expressed as:
\begin{equation}
\begin{aligned}
\left[\bm{u}\right]
=
\begin{bmatrix}
\bm{u}_R & \bm{u}_S &\bm{u}_T
\end{bmatrix},\;\;
\left[\bm{i}\right]
=
\begin{bmatrix}
\bm{i}_R & \bm{i}_S &\bm{i}_T
\end{bmatrix}
\end{aligned}
\end{equation}
so that the geometric power would be calculated as
%
\begin{equation}
\bm{M}
=
\left[\bm{u}\right]\left[\bm{i}\right]^T
\end{equation}
The development of this theory for three-phase and multi-phase systems is of interest for further research.
%
\vspace{-0.15cm}
\section{Conclusion}
In this paper, an improved version of the power theory based on GA has been presented, and it has been named GAPoT.
%
%
%
First, the main shortcomings of current power theories based on GA were identified.
It was shown that the use of $k$-vectors as a basis for the geometric space leads to an unclear definition of the apparent power. Moreover, the energy conservation principle can not be fulfilled without factor correction.
%
%
As an alternative, a transformation that simplifies power definitions and provides a clear meaning to harmonic power has been presented. Also, it is in good agreement with the traditional definition of apparent power based on the product of RMS voltage and current.
%
Current decomposition for load compensation purposes can be easily carried out. 
Through different examples, it is shown that GAPoT theory is a comprehensive tool for analysing and solving single-phase electrical circuits under distorted conditions.
Finally, suggestions for the extension to multi-phase systems were presented.
%
%
%
This proposal opens up future perspectives for the analysis of electrical circuits in the time domain.
\vspace{-0.25cm}
\ifCLASSOPTIONcaptionsoff
\newpage
\fi

\bibliographystyle{IEEEtran}
\bibliography{mybibfile}

\end{document}